\documentstyle{article}

\input pz.sty
\input epsf.sty

\begin{document}

\PZhead{4}{28}{2008}{28 March}

\PZtitletl{SN 2005\lowercase{kd}: another very luminous,}{slowly 
declining type II\lowercase{n} supernova}

\PZauth{D. Yu. Tsvetkov}
\PZinst{Sternberg Astronomical Institute, University Ave.13,
119992 Moscow, Russia; e-mail: tsvetkov@sai.msu.su}

\begin{abstract}
CCD $UBVRI$ photometry is presented for type IIn SN 2005kd.  
The maximum luminosity exceeded $M_V=-19.8$, and SN remained
brighter than $-18$ mag for about 400 days. While overall
photometric evolution is quite similar to SN 1997cy, SN 2005kd
shows a plateau at phases between 119 and 311 days past explosion,
which is a unique feature for SN IIn.  
\end{abstract}

SN 2005kd was discovered by T.Puckett and A.Pelloni with 0.35-m
automated supernova patrol telescope on November 12.22 UT at 
magnitude 17.0, while on November 9 the object was not detected
(fainter than 20 mag). It is located at
$\alpha =4\hr03\mm16\sec.88, \delta =+71\deg 43\arcm 18\arcs.9$
(2000.0), which is $0\arcs.1$ west and $5\arcs$ north from the
center of Sc galaxy PGC 14370 (Puckett, Pelloni, 2005).

SN 2005kd has been found by an Ohio State University group  
(Prieto, 2005) to be a
young type-IIn supernova from a spectrogram (range 390-730 nm)
taken on November 13.3 UT with the MDM 2.4-m telescope; the
spectrum shows a blue continuum and strong hydrogen Balmer and He I lines 
in emission.

\medskip

We started photometric observations of SN 2005kd immediately after
discovery, on 2005 November 13, and continued until 2007 April 16.
Observations were carried out with the following telescopes and
CCD cameras: 60-cm reflector of 
Crimean Observatory of Sternberg Astronomical Institute (C60) 
equipped with Apogee AP-47p camera; 50/70-cm meniscus telescope
of Crimean Observatory (C50) with Meade Pictor 416XT camera; 70-cm 
reflector in Moscow (M70) with Apogee AP-47p (a) or AP-7p (b)
cameras. 
 
The color terms for C60 and M70
were reported by Tsvetkov et al. (2006). The observations at C50
were carried out only with $V$ filter which 
was close to standard system, and no correction was applied.

All image reductions and photometry were made using IRAF.\PZfm 
\PZfoot{IRAF is distributed by the National Optical Astronomy Observatory,
which is operated by AURA under cooperative agreement with the
National Science Foundation}
The position of SN 2005kd is quite close to the center of the host
galaxy, and the subtraction of galaxy background is necessary for 
reliable photometry. The template images were constructed from
frames obtained on 2007 August 8 and 2007 September 25, when SN
was no longer visible. After template subtraction the magnitudes of
SN were derived by PSF fitting relative to a sequence of local standard
stars. The image of SN 2005kd with local standard stars is shown in
Figure 1, and the magnitudes of these stars are reported in Table 1.

\PZfig{12cm}{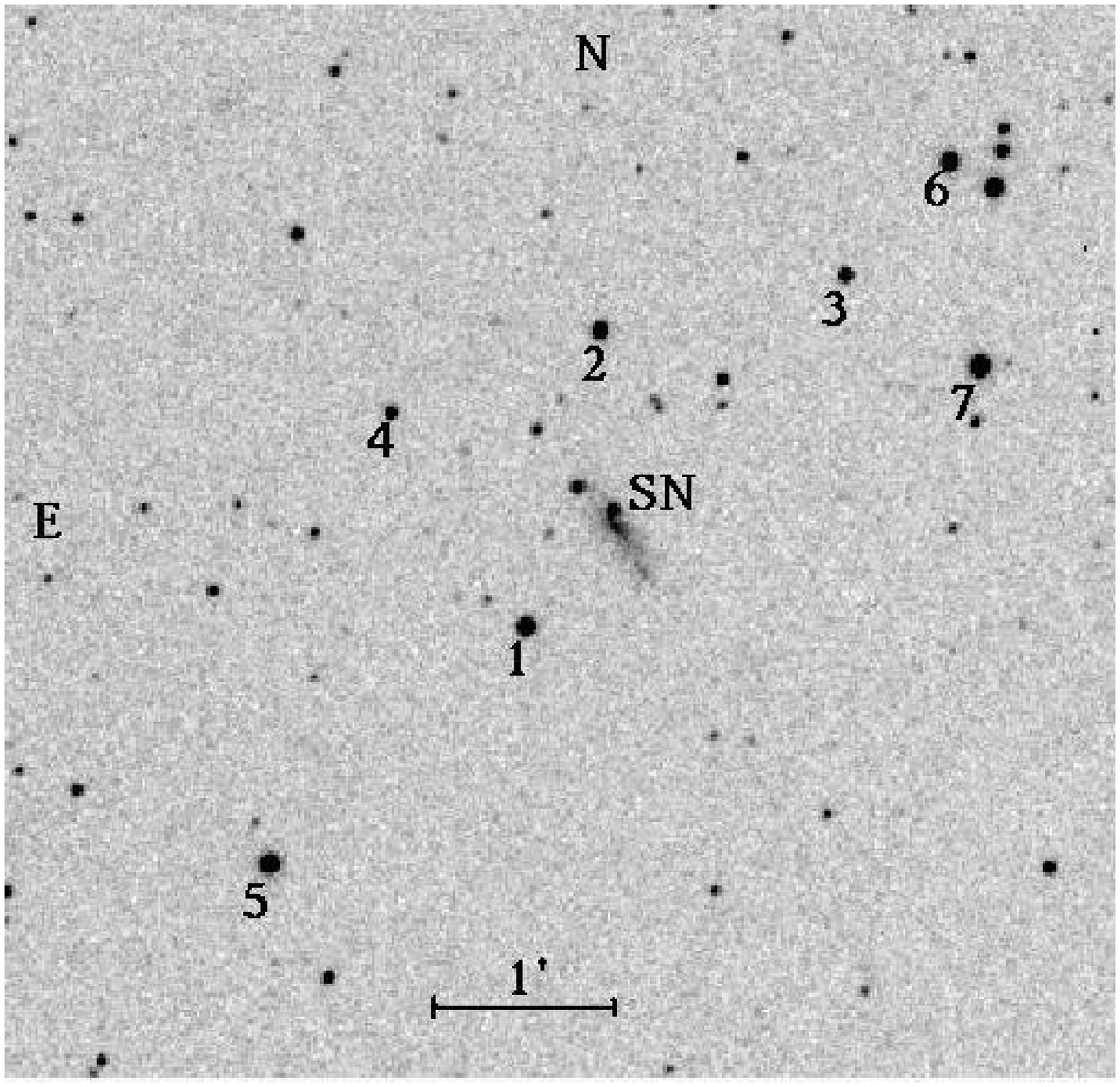}{SN 2005kd in PGC 14370 with local standard
stars}

\begin{table}
\caption{Magnitudes of local standard stars}\vskip2mm
\begin{tabular}{ccccccccccc}
\hline
Star &$U$ & $\sigma_U$ & $B$ & $\sigma_B$ & $V$ & $\sigma_V$ & $R$ &
$\sigma_R$ & $I$ & $\sigma_I$
\\
\hline
1& 15.50& 0.08& 15.21& 0.02& 14.51& 0.03& 14.09& 0.02& 13.68& 0.02 \\ 
2&      &     & 16.15& 0.03& 15.09& 0.02& 14.45& 0.02& 13.92& 0.02 \\
3&      &     & 16.50& 0.05& 15.90& 0.04& 15.41& 0.03& 15.01& 0.02\\
4&      &     & 16.80& 0.04& 16.13& 0.04& 15.62& 0.03& 15.24& 0.02\\
5& 15.76& 0.10& 15.10& 0.03& 14.13& 0.02& 13.62& 0.01& 13.10& 0.02\\
6& 15.36& 0.14& 15.18& 0.05& 14.42& 0.04& 13.98& 0.02& 13.56& 0.02\\
7& 15.33& 0.16& 14.80& 0.05& 13.82& 0.04& 13.32& 0.05& 12.81& 0.03\\ 
\hline
\end{tabular}
\end{table}

The observations of SN 2005kd are presented in Table 2, and the light
curves are shown in Figure 2. Some unfiltered CCD magnitudes were 
reported at IAU Circulars and at SNWeb,\PZfm
\PZfoot{http://www.astrosurf.com/snweb2/2005/05kd/05kdMeas.htm}
they are also plotted in Figure 2. 
The data shows that SN 2005kd was discovered immediately after 
explosion, and our first two observations were on the rising branch
of the light curve. The rate of brightness increase is about 
0.3-0.4 mag day$^{-1}$ in all bands. 
The outburst most likely occurred on 2005 November 10 or 11,
and we accept JD 2453685 as the date of explosion. 
Unfortunately, we missed the most
interesting part of the light curve and cannot reliably establish
the shape of the light curve peak and the maximum luminosity.
Our next observation was only on 2006 January 3, and on this date
SN was brightest of all our data set. The magnitudes from SNWeb
allow to suggest quite flat maximum, but they also have a large gap.
After small drop from the maximum the SN entered a plateau stage, 
which lasted for at least 192 days, from day 119 until day 311 past
explosion. Another gap in observations does not allow to determine
the length of the plateau more definitely. Since 2005 September 17
(day 341) until the end of our observations at day 522 the SN
is gradually fading, but at different rates. Until day 405
the decline is slow, with rates 0.0077 mag day$^{-1}$ in $R$ and $I$ bands,
0.0087 mag day$^{-1}$ in $V$ and 0.013 mag day$^{-1}$ in $B$.
The late decline is about two times faster: 0.017 mag day$^{-1}$ in 
$R$ and $I$ bands, 0.025 mag day$^{-1}$ in $B$ and $V$.

\begin{table}
\caption{Observations of SN 2004A}\vskip2mm
\begin{tabular}{cccccccccccl}
\hline
JD 2450000+ & $U$ & $\sigma_U$ & 
$B$ & $\sigma_B$ & $V$ & $\sigma_V$ & $R$ & $\sigma_R$ &
$I$ & $\sigma_I$ & Tel.\\
\hline
3688.47& 15.38& 0.12& 16.06& 0.03&  16.06& 0.03 & 15.77& 0.03 & 15.77& 0.06& C60\\
3689.52& 15.15& 0.09& 15.72& 0.03&  15.68& 0.02 & 15.48& 0.02 & 15.41& 0.03& C60\\
3739.47&      &     & 15.67& 0.03&  15.10& 0.03 & 14.59& 0.02 & 14.17& 0.03& M70b\\
3804.36&      &     & 16.12& 0.03&  15.54& 0.03 & 14.91& 0.02 & 14.47& 0.02& M70a\\
3822.30&      &     & 16.06& 0.03&  15.60& 0.04 & 14.94& 0.02 & 14.51& 0.03& M70a\\
3831.27&      &     & 16.07& 0.04&  15.59& 0.03 & 14.94& 0.02 & 14.49& 0.02& M70a\\
3852.28&      &     & 16.02& 0.03&  15.66& 0.03 & 14.88& 0.02 & 14.50& 0.02& M70a\\
3872.34&      &     & 16.03& 0.03&  15.59& 0.02 & 14.86& 0.02 & 14.44& 0.03& M70a\\
3996.42&      &     & 16.42& 0.07&  16.61& 0.04 & 15.53& 0.02 & 15.20& 0.03& M70b\\
4026.44& 16.93& 0.14& 17.00& 0.04&  16.86& 0.03 & 15.74& 0.03 & 15.35& 0.07& M70b\\ 
4044.61&      &     &      &     &  16.93& 0.07 & 16.08& 0.04 &      &     & C60\\
4056.49&      &     &      &     &  17.07& 0.08 &      &      &      &     & C50\\
4059.38&      &     &      &     &  17.26& 0.04 &      &      &      &     & C50\\
4059.53&      &     & 17.57& 0.03&  17.38& 0.04 & 15.97& 0.06 & 15.70& 0.08& C60\\
4062.38&      &     &      &     &  17.26& 0.03 &      &      &      &     & C50\\
4090.48&      &     & 17.69& 0.04&  17.21& 0.06 & 16.07& 0.04 & 15.81& 0.03& M70b\\
4118.30&      &     & 18.50& 0.04&  18.17& 0.07 & 16.85& 0.03 & 16.36& 0.04& M70b\\
4127.25&      &     &      &     &       &      & 17.04& 0.08 &      &     & M70b\\
4131.27&      &     & 18.99& 0.12&       &      & 17.06& 0.05 & 16.81& 0.09& M70b\\
4143.23&      &     & 18.94& 0.10&  19.07& 0.20 & 17.35& 0.06 & 16.83& 0.08& M70b\\
4158.26&      &     &      &     &       &      & 17.70& 0.05 & 17.21& 0.12& M70b\\
4180.30&      &     &      &     &       &      & 17.99& 0.06 & 17.40& 0.06& M70b\\
4183.28&      &     & 19.78& 0.13&  19.44& 0.28 & 17.70& 0.09 & 17.29& 0.08& M70b\\
4187.27&      &     &      &     &       &      & 17.80& 0.04 & 17.30& 0.04& M70b\\
4201.27&      &     &      &     &       &      &18.21 &0.10  &17.58 &0.08 & M70b\\
4207.27&      &     &      &     &       &      & 18.11& 0.04 & 17.80& 0.10& M70b\\
\hline
\end{tabular}
\end{table}

\PZfig{12cm}{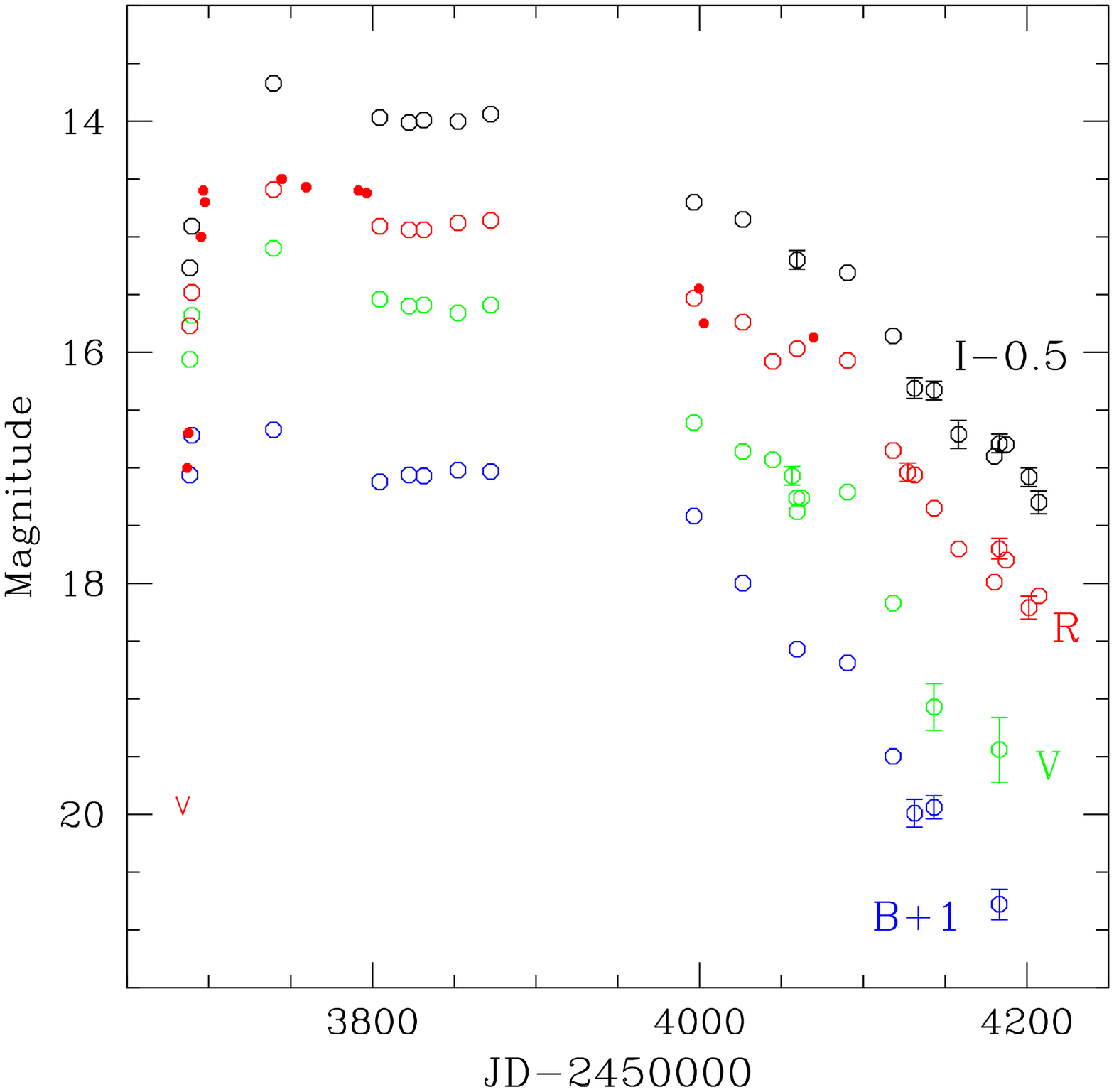}{$BVRI$ light curves of SN 2005kd,
showing our photometry (circles)
and the magnitudes reported at SNWeb (dots). 
Error bars for our magnitudes are plotted only when they
exceed the size of a point}

\PZfig{12cm}{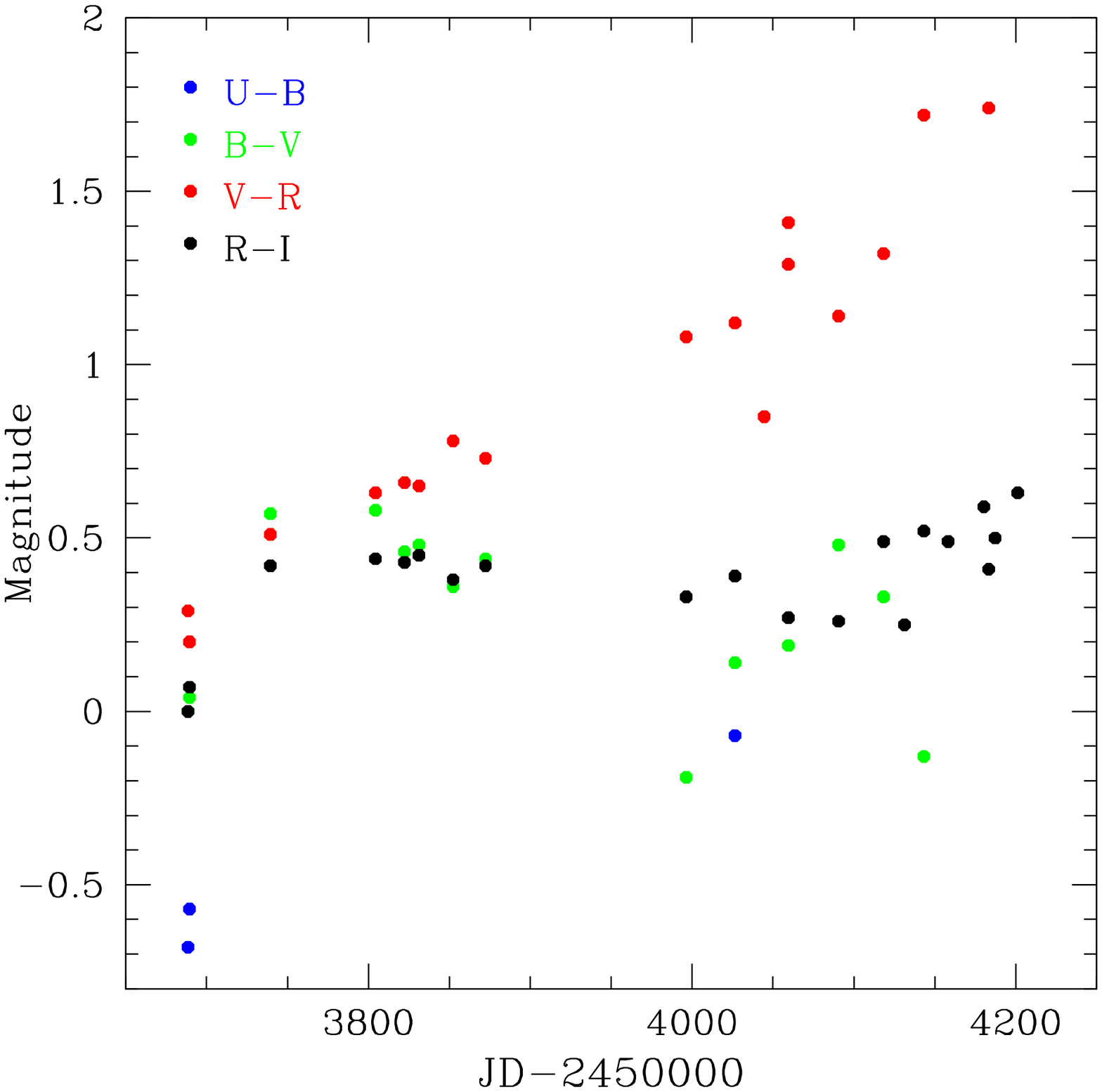}{Color curves for SN 2005kd}

The color curves are presented in Figure 3. $(V-R)$ color is
gradually increasing, $(R-I)$ remains nearly constant after
initial increase, while $(B-V)$ reaches maximum on the 
plateau and then slightly decreases. 

If we take for PGC 14370 distance modulus $\mu=34.07$ and Galactic 
extinction $A_V=0.87$ from NED,\PZfm
\PZfoot{http://nedwww.ipac.caltech.edu}
then the absolute magnitude on 2006 January 3 (day 54) is
$M_V=-19.84$. The real maximum luminosity can be significantly
higher, because we missed the peak of the light curve and 
do not know the extinction in the host galaxy.    

The absolute $V$ light curve of SN 2005kd is shown in Figure 4 and
compared to the light curves of well-studied slowly declining 
type IIn SNe: 1997cy (Germany et al., 2000), 
1999E (Rigon et al., 2003), 
1995G (Pastorello et al., 2002), 1988Z (Turatto et al., 1993).
The similarity of overall photometric evolution of SNe 2005kd and
1997cy is evident, although the differences are also noticeable.
We note that for both SNe the rate of brightness
decline changed at about the same phase, close to day 400. 
SN 1999E is fainter, but the shape of the light curve is the same
as for SN 1997cy; SNe 1988Z and 1995G are much fainter and have
different light curves. The plateau of SN 2005kd is the unique
feature for SNeIIn, nothing similar can be seen on all other light
curves.

\PZfig{12cm}{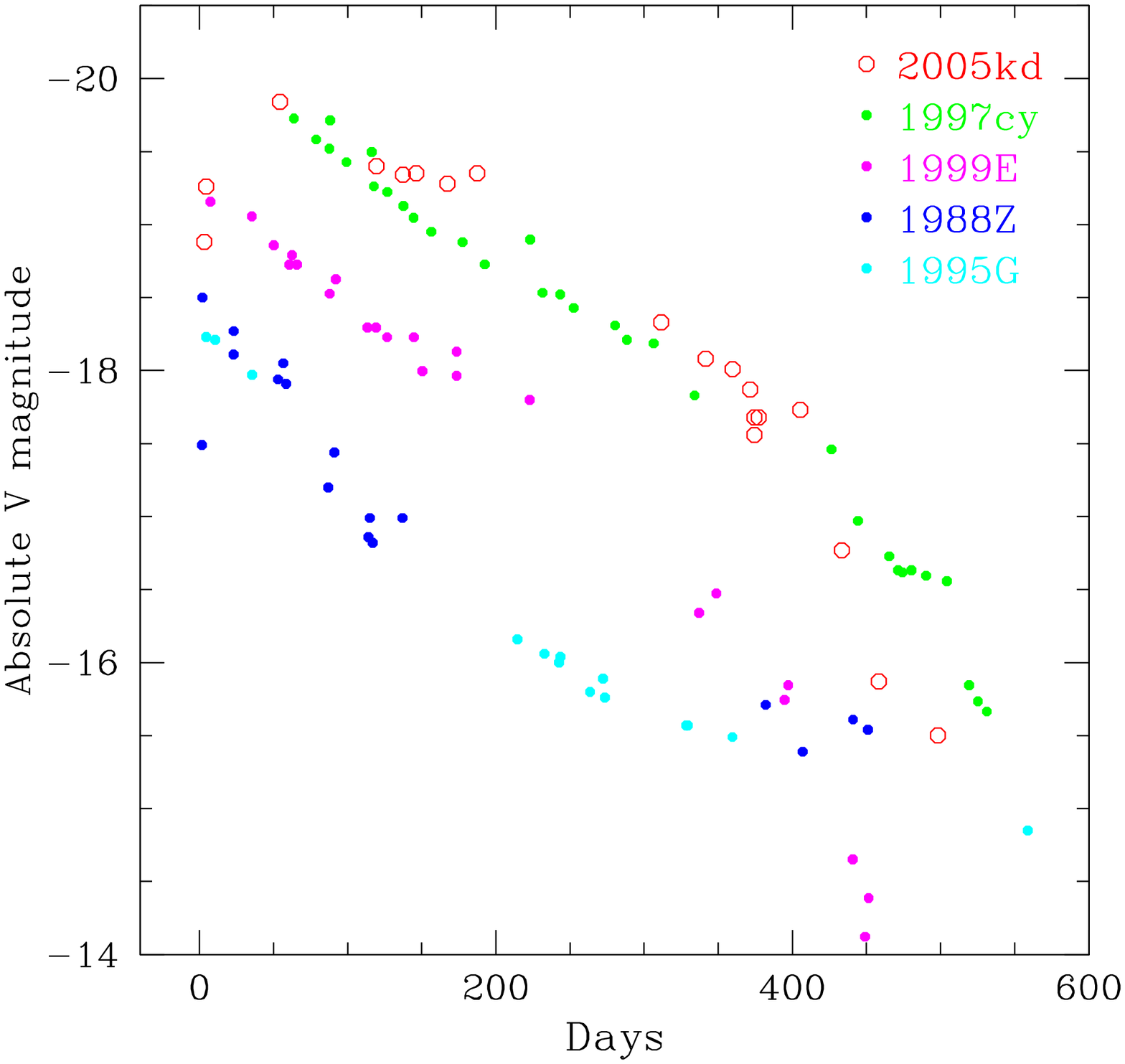}{The absolute $V$ light curve of SN 2005kd
compared to those for SNe IIn 1997cy, 1999E, 1995G and 1988Z}

The $(B-V)_0$ color curves for these SNe are compared in Figure 5.
SNe 1997cy, 1999E and 1999G have similar color evolution, after initial
reddening $(B-V)$ color remains nearly constant, at about 0.5-0.6 mag.
SNe 2005kd and 1988Z are certainly bluer, and at the phases greater 
then 300 days SN 2005kd is the bluest among these objects.

\PZfig{12cm}{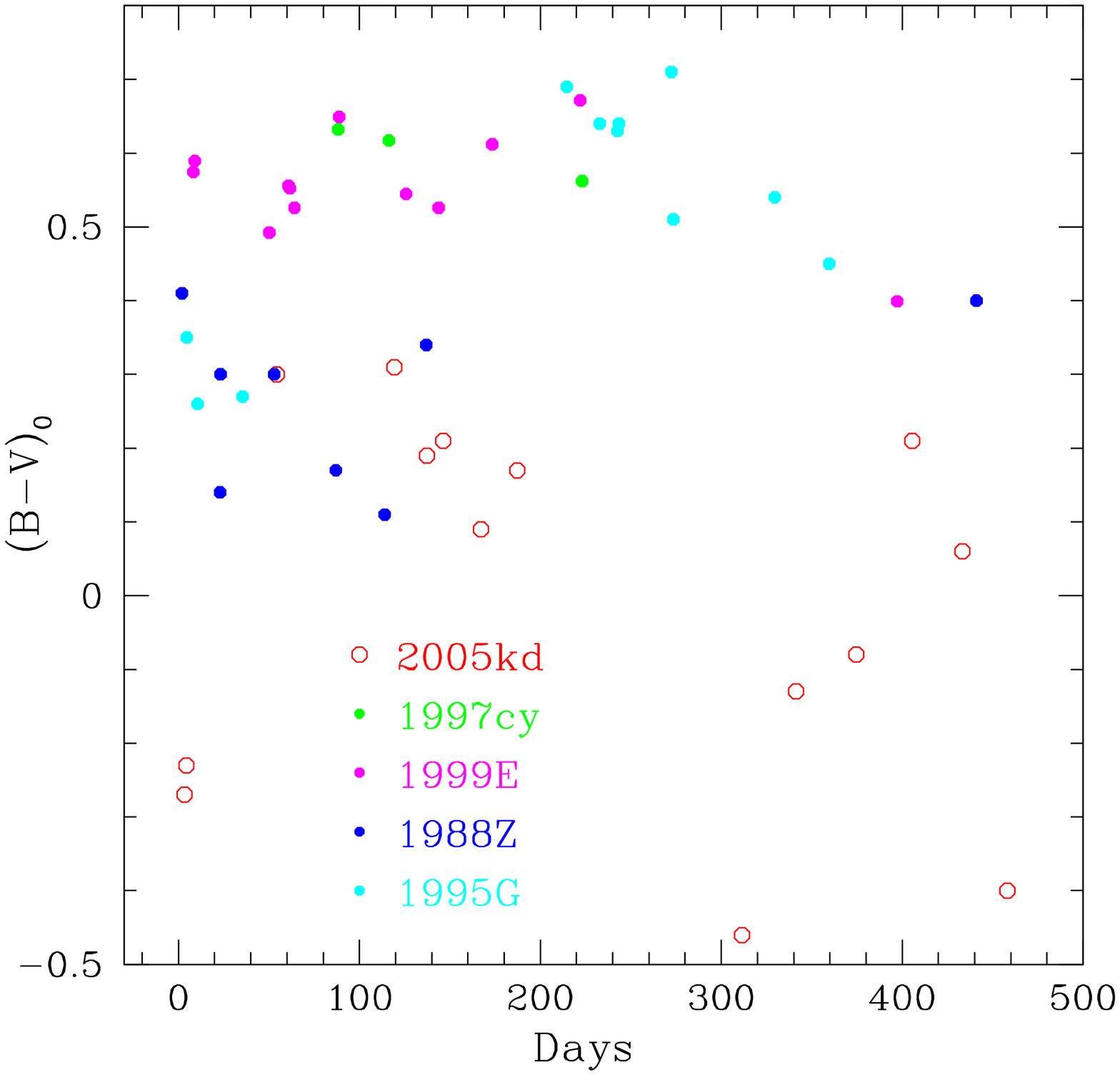}{The $(B-V)_0$ color curve for SN 2005kd
compared to those for SNe IIn 1997cy, 1999E, 1995G and 1988Z}

We can roughly estimate the lower limit to the energy radiated by 
SN 2005kd in 
$UBVRI$ bands during first 500 days of evolution. Assuming the
distance and extinction reported earlier and using simple linear
interpolations in the gaps, we obtain $E_{rad}=3.2\cdot10^{50}$ ergs.

SN 2005kd is among the most luminous SNe ever observed. As for
other SNe IIn, its high energy release is likely due to the 
interaction of ejecta with a dense circumstellar medium. For all
well-studied SNe IIn the brightness decline was slow but 
gradual, without periods of constant luminosity. The plateau
lasting at least 192 days observed for SN 2005kd is a unique
feature for SNe IIn. Unfortunately, the gaps in our data does not
allow to trace the peak of the light curve and the end of plateau.

\medskip

This research has made use of the NASA/IPAC Extragalactic Database
(NED) which is operated by the Jet Propulsion Laboratory, California
Institute of Technology, under contract with NASA.

\newpage
\references
Germany, L.M., Reiss, D.J., Sadler, E.M., Schmidt, B.P., Stubbs, C.W.,
2000, {\it Astrophys. J.}, {\bf 533}, 320

Pastorello, A., Turatto, M., Benetti, S., Cappellaro, E.,
Danziger, I.J., Mazzali, P.A., Patat, F., Filippenko, A.V., 
Schlegel, D.J., Matheson, T., 2002, {\it MNRAS}, {\bf 330}, 844 

Prieto, J., 2005, {\it IAU Circ.}, No. 8630

Puckett, T., Pelloni, A., 2005, {\it IAU Circ.}, No. 8630 

Rigon, L., Turatto, M., Benetti, S., Pastorello, A., Cappellaro, E.,
Aretxaga, I., Vega, O., Chavushyan, V., Patat, F., Danziger, I.J., Salvo,
M., 2003, {\it MNRAS}, {\bf 340}, 191 

Tsvetkov, D.Yu., Volnova, A.A., Shulga, A.P., Korotkiy, S.A., 
Elmhamdi, A., Danziger, I.J., Ereshko, M.V., 2006,
{\it Astron. Astrophys.}, {\bf 460}, 769

Turatto, M., Cappellaro, E., Danziger, I.J., Benetti, S., 
Gouiffes, C., Della Valle, M., 1993, {\it MNRAS}, {\bf 262}, 128 

\endreferences
\end{document}